\documentclass{PoS}

\title{Light and strange quark masses from $N_f=2+1$ simulations with Wilson fermions}

\ShortTitle{Light and strange quark masses}

\author{M. Bruno$^a$, I. Campos$^b$,
     \speaker{J. Koponen}$^c$,
     C. Pena$^e$, D. Preti$^f$,
     A. Ramos$^g$ and A. Vladikas$^c$ for the ALPHA Collaboration\\
     \llap{$^a$} Theoretical Physics Department, CERN, CH-1211 Geneva 23, Switzerland\\
     \llap{$^b$} Instituto de F\'{i}sica de Cantabria IFCA-CSIC, \\
                ~Av. de los Castros, 39005 Santander, Cantabria, Spain\\
     \llap{$^c$} INFN sezione di Roma tor Vergata, Via della Ricerca Scientifica 1, I-00133 Rome, Italy\\
     \llap{$^e$} Departamento de F\'{i}sica Te\'orica and Instituto de F\'{i}sica Te\'orica UAM-CSIC\\
                ~Universidad Aut\'onoma de Madrid, E-28049 Madrid, Spain\\
     \llap{$^f$} INFN sezione di Torino, Via Pietro Giuria 1, I-10125 Turin, Italy\\
     \llap{$^g$} School of Mathematics, Trinity College Dublin, Dublin 2, Ireland\\
     E-mail:  \email{jonna.koponen@roma2.infn.it}
}

\abstract{
We present a nearly final analysis of the $u/d$ and $s$ quark
masses, extracted using the PCAC quark masses reported in \cite{Bruno:2016plf}.
The data is based on the CLS $N_f = 2 + 1$ simulations with Wilson/Clover quarks
and L\"uscher-Weisz gauge action, at four $\beta$ values (i.e. lattice spacings)
and a range of quark masses. We use the ALPHA results of \cite{Campos:2018ahf} 
for non-perturbative quark mass renormalisation and RG-running from hadronic 
to electroweak scales in the Schr\"odinger Functional scheme. Quark masses are quoted
both in the $\overline{\rm MS}$ scheme and as RGI quantities.          
}

\FullConference{The 36th Annual International Symposium on Lattice Field Theory - LATTICE2018\\
		22-28 July, 2018\\
		Michigan State University, East Lansing, Michigan, USA.}

\usepackage{amsmath}
\DeclareMathOperator{\Tr}{Tr}
\DeclareMathOperator{\tr}{tr}

\begin{document}

\section{Introduction and setup}
\label{sec:intro}

Here we report of the ALPHA Collaboration's analysis of the $u/d$ and $s$ quark masses using Wilson
fermions. The starting point is the computation of light PCAC bare quark masses (up/down and strange),
performed in refs.~\cite{Bruno:2016plf,Bruno:2014jqa} with $N_f = 2+1$ dynamical sea quarks. The gauge
action is the L\"uscher-Weisz action with tree-level coefficients~\cite{Luscher:1984xn}, and the
fermion action is non-perturbatively $\mathcal{O}(a)$ improved, with the value of the clover
coefficient $c_{\rm\scriptscriptstyle SW}$ obtained in~\cite{Bulava:2013cta}. The boundary conditions
are periodic in space and open in time, as detailed in ref.~\cite{Luscher:2012av}.

Simulations have been carried out at four lattice spacings: $a\approx 0.050$, $0.064$, $0.076$
and $0.086$~fm, corresponding to lattice couplings $\beta=3.7$, $3.55$, $3.46$ and $3.40$
respectively. In order to keep finite-size effects under control, all ensembles have
$LM_{\pi} \gtrsim 4$ and the time extent varies from $T=2L$ to $T=3L$ (where $L^3\times T$ is
the lattice size). The pion mass $M_{\pi}$ varies between $200$~MeV and $420$~MeV, and the kaon mass
$M_{\rm K}$ between $420$~MeV and $470$~MeV. A detailed overview of the simulations
can be found in ref.~\cite{Bruno:2016plf}.

For each lattice coupling $\beta$ we have ensembles with different values of the hopping
parameters $\kappa_1 = \kappa_2$ and $\kappa_3$ (except for $\beta = 3.46$ where we only have one
ensemble with three degenerate quark masses). The bare subtracted quark masses are defined as
$m_{{\rm q},r} = 1/(2\kappa_r) - 1/(2\kappa_{\rm crit})$, with $\kappa_{\rm crit}$ the critical (chiral)
point. The index $r$ labels quark flavours: we use values 1 and 2 for the two degenerate light quark
flavours ($u$ and $d$), and 3 for the strange quark. These masses are chosen so that their mass matrix,
at a given $\beta$, satisfies the condition
\begin{equation}
\Tr M_{\rm q}  = 2m_{{\rm q},1}+m_{{\rm q},3}={\rm constant}.
\end{equation}
This condition ensures that the improved bare gauge coupling
\begin{equation}
\label{eq:trM0}
\tilde g_0^2 \equiv g_0^2 \Big(1+\dfrac{1}{N_f} b_g a \Tr M_{\rm q} \Big)
\end{equation}
is constant up to $\mathcal{O}(a^2)$ effects, for any $b_g$.
Consequently, in the improved theory a constant $\tilde g_0^2(a)$ corresponds to fixed lattice
spacing. However, a constant $\Tr M_{\rm q}$ does not correspond to a constant trace of the
renormalised quark mass matrix $M^{\rm R}$,
since~\cite{luscher:1996sc,Bhattacharya:2005rb}
\begin{equation}
\Tr M^{\rm R}  = Z_m[(1+a\bar d_m \Tr M_{\rm q}) \Tr M_{\rm q} + ad_m \Tr(M_{\rm q}^2)] +\mathcal{O}(a^2).
\end{equation}
The $d_m$ counter-term is proportional to squared masses and this violates a constant $\Tr M^{\rm R}$
requirement by $\mathcal{O}(a)$ effects. This is an undesirable feature, since we wish to stay on a
constant-physics trajectory (up to $\mathcal{O}(a^2)$), as the bare parameters (masses) are varied.
This problem can be be avoided
by redefining the chiral trajectory in terms of $\phi_4 =$~const.~\cite{Bruno:2016plf}, where
\begin{equation}
\label{eq:defphi4}
\phi_4  \equiv  8t_0\Big( M_{\rm K}^2+\dfrac{1}{2} M_{\pi}^2 \Big).
\end{equation}
Here $t_0 $ is a gluonic dimension-two quantity defined using the Wilson flow~\cite{Luscher:2010iy}.
This requirement gives a Symanzik-improved constant physics condition. The improved bare coupling
$\tilde g_0^2$ now suffers from $\mathcal{O}(a m_q \tr M_q )$ discretisation effects due to
higher-order $\chi$PT contributions, but these turn out to be small and can be ignored.

The values of the bare quark masses are chosen so that one is approximately at the physical value of
$\phi_4$. The precise value of $\phi_4^{\rm phys}$ (and $t_0^{\rm phys}$) is found a posteriori, as part
of the scale setting. The small differences between the target and measured values of $\phi_4$ and
$\phi_2$ for each ensemble can be corrected for by expanding observables in powers of $\Delta m_q$,
and computing the relevant coefficients --- see ~\cite{Bruno:2016plf} for details. 
The aim is to express the computed quantities of interest (in our case the quark masses) as functions of 
\begin{equation}
\phi_2 \equiv 8t_0M_{\pi}^2,
\label{eq:defphi2}
\end{equation}
with $\phi_4$ held fixed at $\phi_4^{\rm phys}$, and eventually extrapolate them to
$\phi_2^{\rm phys} = 8t_0^{\rm phys} m_{\pi}^2$ (where $m_\pi$ is the physical pion mass).

Following ref.~\cite{Bruno:2016plf} we define the bare correlation functions
\begin{equation}
  f_{\rm P}^{rs}(x_0,y_0) =  -\dfrac{a^6}{L^3}\sum_{\vec x,\vec y}
  \langle P^{rs}(x_0,\vec x)P^{sr}(y_0,\vec y)\rangle ,\quad
  f_{\rm A}^{rs}(x_0,y_0)  =  -\dfrac{a^6}{L^3}\sum_{\vec x,\vec y}
  \langle A_0^{rs}(x_0,\vec x)P^{sr}(y_0,\vec y)\rangle ,
\end{equation}
where the bare pseudoscalar density and  axial current are
\begin{equation}
P^{rs}(x) = \bar\psi^r(x) \gamma_5 \psi^s(x), \quad
A_0^{rs}(x) = \bar\psi^r(x) \gamma_0 \gamma_5 \psi^s(x) + ac_{\rm\scriptscriptstyle A} \partial_0 P^{rs}(x)
\end{equation}
(the indices $r,s$ label quark flavours). These two-point functions are estimated with stochastic
sources located near the boundaries as described in refs.~\cite{Bruno:2016plf,Bruno:2014jqa}.
The $\mathcal{O}(a)$-improvement coefficient $c_{\rm\scriptscriptstyle A}$ has been tuned non-perturbatively in
ref.~\cite{Bulava:2015bxa}. The bare PCAC mass is then defined through the ratio
\begin{equation}
\label{eq:PCACm}
m_{rs}  =  \dfrac{f_{\rm A}^{rs}(x_0+a,y_0)-f_{\rm A}^{rs}(x_0-a,y_0)}{4 f_{\rm P}^{rs}(x_0,y_0)}.
\end{equation}

This PCAC quark mass is re-expressed as a dimensionless quantity through the definition
\begin{equation}
\label{eq:defphiPCAC}
\phi_{rs}  \equiv \sqrt{8 t_0} m_{rs},
\end{equation}
and the corresponding renormalisation group invariant (RGI) dimensionless quantity is then given by
\begin{equation}
\label{eq:phi-rs-RGI}
\phi^{\rm RGI}_{rs}  = \dfrac{M}{\overline{m}(\mu_{\rm had})} \phi^{\rm R}(\mu_{\rm had})
= Z_{\rm M} \Big( 1 + (\tilde{b}_{\rm\scriptscriptstyle A}-\tilde{b}_{\rm\scriptscriptstyle P}) am_{rs} +
(\bar{b}_{\rm\scriptscriptstyle A}-\bar{b}_{\rm\scriptscriptstyle P})a\Tr M_{\rm q} \Big)\phi_{rs}
   +  \mathcal{O}(a^2)  ,
\end{equation}
where the renormalisation coefficient $Z_{\rm M}$ is
\begin{equation}
Z_{\rm M}(g_0^2)=\dfrac{M}{\overline{m}(\mu_{\textrm{had}})}\dfrac{Z_{\rm A}(g_0^2)}{Z_{\rm P}(g_0^2,a\mu_{\textrm{had}})}.
\end{equation}
The first factor, $M/\overline{m}(\mu_{\textrm{had}})$, is the ratio of the RGI quark mass $M$ (in
physical units) to the renormalised quark mass $\overline{m}(\mu_{\textrm{had}})$. The second factor,
$Z_{\rm A}(g_0^2)/Z_{\rm P}(g_0^2,a\mu_{\textrm{had}})$, is the ratio of the axial current normalisation
$Z_{\rm A}(g_0^2)$ to the pseudoscalar density $Z_{\rm P}(g_0^2,a\mu_{\textrm{had}})$. The former is scale
independent and depends solely on the bare gauge coupling, while the latter depends on a renormalisation
scheme and a renormalisation scale which we set in the hadronic region of low energies.

The quark mass RG-running was carried out non-perturbatively up to $\mu_{\textrm{pt}} \sim 100$~GeV
in the Schr\"odinger Functional scheme for a theory with $N_f=3$ massless quarks in
ref.~\cite{Campos:2018ahf}. Standard step-scaling functions were obtained in the continuum,
by extrapolating results computed on small lattices at fixed renormalisation scale $\mu$.
Beyond the scale $\mu_{\textrm{pt}}$ the RG-running is done perturbatively
(at 2-loops for the quark mass and 3-loops for the gauge coupling). The result quoted in
ref.~\cite{Campos:2018ahf} is
\begin{equation}
\label{eq:ZMovermR}
\dfrac{M}{\overline{m}(\mu_{\textrm{had}})}=0.9148(88)  ,
\end{equation}
for $\mu_{\rm had} = 233(8)~{\rm MeV}$; the error encompasses both statistical and systematic effects.

We use the axial current renormalisation parameter $Z_{\rm A}(g_0^2)$ obtained on the chirally rotated
Schr\"odinger Functional setup in ref.~\cite{DallaBrida:2018tpn}.
The renormalisation parameter $Z_{\rm P}(g_0^2,\mu_{\rm had})$ was computed in ref.~\cite{Campos:2018ahf} for
the same action and in the Schr\"odinger Functional scheme at fixed scale $\mu_{\rm had}$, in the range of
inverse gauge couplings $\beta \in [3.40,3.85]$. This is the range covered by the large volume ensembles of
ref.~\cite{Bruno:2016plf}, from which our bare PCAC masses $\phi_{rs}$ are extracted. The final result is
summarised as
\begin{equation}
\label{eq:ZM-beta}
Z_{\rm M}(g_0)=Z_{\rm M}^{(0)}+Z_{\rm M}^{(1)}(\beta - 3.79)+Z_{\rm M}^{(2)}(\beta - 3.79)^2,
\end{equation}
where
\begin{equation}
\label{eq:ZM-012}
  Z_{\rm M}^{(0)}= 2.270073\times \dfrac{M}{\overline{m}(\mu_{\textrm{had}})},\,\,
  Z_{\rm M}^{(1)}= 0.125658\times \dfrac{M}{\overline{m}(\mu_{\textrm{had}})},\,\, 
  Z_{\rm M}^{(2)}=-0.464575\times \dfrac{M}{\overline{m}(\mu_{\textrm{had}})},
\end{equation}
with covariance matrix
\begin{equation}
\label{eq:ZM-cov}
\textrm{cov}(Z_{\rm M}^{(i)},Z_{\rm M}^{(j)})=
\begin{pmatrix}
\begin{array}{rrr}
 0.164635\times 10^{-4} & 0.215658\times 10^{-4} & -0.754203\times 10^{-4} \\
 0.215658\times 10^{-4} & 0.121072\times 10^{-2} &  0.308890\times 10^{-2} \\
-0.754203\times 10^{-4} & 0.308890\times 10^{-2} &  0.953843\times 10^{-2} 
\end{array}
\end{pmatrix}   .
\end{equation}
The quoted errors only contain the uncertainties from the determination of $Z_{\rm A}$ and $Z_{\rm P}$ at
the hadronic scale.  The error of the total running factor $M/\overline{m}(\mu_{\textrm{had}})$ in
eq.~(\ref{eq:ZMovermR}) only affects the continuum limit, and is only included after the extrapolation
to vanishing lattice spacing.

As seen from equation \eqref{eq:phi-rs-RGI}, besides the renormalisation parameter $Z_{\rm M}$, we
also need the mass-dependent improvement coefficients
$(\tilde{b}_{\rm\scriptscriptstyle A}-\tilde{b}_{\rm\scriptscriptstyle P})$ and
$(\bar{b}_{\rm\scriptscriptstyle A}-\bar{b}_{\rm\scriptscriptstyle P})$.
We ignore the latter, as it is $\mathcal{O}(g_0^4)$ in perturbation
theory. The coefficient $(\tilde{b}_{\rm\scriptscriptstyle A}-\tilde{b}_{\rm\scriptscriptstyle P})$
is known in 1-loop perturbation theory
and has the value $(\tilde{b}_{\rm\scriptscriptstyle A}-\tilde{b}_{\rm\scriptscriptstyle P}) = -0.0012 g_0^2$.
However, we use a preliminary
value from a non-perturbative determination from the ALPHA Collaboration~\cite{CarlKoster}:
\begin{equation}
  \tilde{b}_{\rm\scriptscriptstyle A}-\tilde{b}_{\rm\scriptscriptstyle P}
  =\dfrac{1.03652(g_0^2)^3-0.863388(g_0^2)^4+0.109868(g_0^2)^5}{0.956281(g_0^2)^3-0.246337(g_0^2)^4-0.116847(g_0^2)^5}.
\end{equation}
The perturbative and non-perturbative values are quite different for the ensembles used in this study,
but the effect on the final result is negligible.

\section{Chiral Fits}

Having obtained the renormalised, dimensionless quantities $\phi^{\rm R}_{12}$ and
$\phi^{\rm R}_{13}$ (where indices 1,2 refer to the degenerate light quarks and 3 is the
strange quark at the physical point) we now proceed
to do the chiral and continuum extrapolation. We adapt the standard $\chi$PT expressions to
our specific parametrisation of our data, which leads to
\begin{align}
\label{eq:combinedfit-0}
\phi^{\rm R}_{12} &= \phi_2\bigg[b_1-b_2\phi_2 -c_{\textrm{log}}K\Big(\bar{L}_{\pi}
-\dfrac{1}{3}\bar{L}_{\eta}\Big)\bigg] + c_{a1} \dfrac{a^2}{8t_0}, \\
\phi^{\rm R}_{13} &= \dfrac{2\phi_4-\phi_2}{2}\bigg[b_1 - b_2 \Big(\dfrac{2\phi_4-\phi_2}{2} \Big)
-c_{\textrm{log}}\dfrac{2}{3}K\bar{L}_{\eta}\bigg] + c_{a2} \dfrac{a^2}{8t_0}. \nonumber
\end{align}
Note that $\phi^{\rm R}_{12}$ and $\phi^{\rm R}_{13}$ are functions of $\phi_2$ only, $\phi_4$ being held
constant. They have common fit parameters $b_1$, $b_2$ and $c_{\textrm{log}}$, arising from  NLO $\chi$PT.
The chiral logs are $\bar{L}_{\pi}=\phi_2\ln{\phi_2}$ and $\bar{L}_{\eta}=\phi_{\eta}\ln{\phi_{\eta}}$,
where $\phi_{\eta}\equiv (4\phi_4-3\phi_2)/3$ and the fit parameters relate to LEC's as
\begin{equation}
b_1=\dfrac{1}{2B_0\sqrt{8t_0}}\bigg[1-\dfrac{32}{8t_0f_0^2}(2L_6-L_4)\phi_4\bigg], \quad
b_2=\dfrac{1}{2B_0\sqrt{8t_0}}\dfrac{16}{8t_0f_0^2}(2L_8-L_5), \quad
c_{\rm log}=\dfrac{1}{2B_0\sqrt{8t_0}}, \nonumber
\end{equation}
and
\begin{equation}
K=\dfrac{1}{16\pi^28t_0f_{\pi\rm K}^2}, \quad f_{\pi\rm K}\equiv \dfrac{2}{3}\Big(f_K+\dfrac{1}{2}f_{\pi}\Big).
\end{equation}
The fit parameters $c_{a1}$, $c_{a2}$ arise from the parametrisation of the discretisation effects in our
Sy\-man\-zik-improved setup. It is implied that the dominant discretisation error is mass-independent;
i.e. corrections of $\mathcal{O}(a^2 \phi_2)$ may be ignored. This is supported by work on Wilson $\chi$PT
\cite{Bar:2003mh,Herdoiza:2013sla}. In some test-fits, where such a mass-dependent term was also allowed, it
turned out to be small.

These chiral formulae can be combined to form the ratio of the two PCAC masses,
\begin{align}
  \dfrac{\phi^{\rm R}_{12}}{2\phi^{\rm R}_{13}}=&\dfrac{\phi_2}{2\phi_4 -\phi_2}\Big[ 1+ \dfrac{b_2}{b_1}\phi_4
    -\dfrac{3b_2}{2b_1}\phi_2 -\dfrac{c_{\textrm{log}}K}{b_1} \Big(\bar{L}_{\pi}-\bar{L}_{\eta}\Big) \Big ]
  + c_{a} \dfrac{a^2}{8t_0}\Big(1-\dfrac{2\phi_2}{2\phi_4 -\phi_2}\Big) \nonumber \\
  \stackrel{LO}{\approx}&\dfrac{\phi_2}{2\phi_4 -\phi_2}\Big[ 1+ \dfrac{b_2}{b_1}\phi_4
    -\dfrac{3b_2}{2b_1}\phi_2 -K\Big(\bar{L}_{\pi}-\bar{L}_{\eta}\Big) \Big ]
  + c_{a} \dfrac{a^2}{8t_0}\Big(1-\dfrac{2\phi_2}{2\phi_4 -\phi_2}\Big).
\end{align}
The ratio has the advantage of cancelling the renormalisation constants. The form of the cutoff effects
has been tailored to satisfy the exact constraint
\begin{equation}
\dfrac{\phi^{\rm R}_{12}}{\phi^{\rm R}_{13}}\bigg|_{m_l=m_h}=1.
\end{equation}
Another combination to study is
\begin{equation}
  4\dfrac{\phi^{\rm R}_{12}}{2\phi_4 -\phi_2}+\dfrac{\phi^{\rm R}_{12}}{\phi_2}=
  3b_1+2b_2\phi_4+c_{\rm log}K\Big(\bar{L}_{\pi}-\bar{L}_{\eta}\Big)+c_a^{\prime}\dfrac{a^2}{8t_0},
\label{eq:pregolden}
\end{equation}
as this will show how sensitive we are to the chiral logarithms.

The analysis is carried out using the library described in \cite{Ramos}, and standard MINPACK
routines are used for $\chi^2$ minimisation. Errors in abscissa variables -- $\phi_2$, $\phi_4$ and
$K$ -- are included in the fit. The error analysis is carried out using the Gamma method
approach and automatic differentiation for error propagation (see \cite{Ramos} and references therein).
This takes into account all existing correlations in the data, and computes autocorrelation functions
(including exponential tails) to estimate the uncertainties correspondingly. Following
\cite{Bruno:2016plf}, the values $\tau_{\rm exp}$ used in the analysis are those quoted in
\cite{Bruno:2014jqa}, namely
\begin{equation}
\tau_{\rm exp} = 14(3)\dfrac{t_0}{a^2}.
\end{equation}
This is a very conservative estimate for our data.

Doing combined fits to various combinations of the ratio $\phi^{\rm R}_{12}/(2\phi^{\rm R}_{13})$,
$\phi^{\rm R}_{13}$, $\phi^{\rm R}_{12}$ and eq.~\eqref{eq:pregolden} shows that our most stable fits are:
\begin{itemize}
\item[fit 1:] combined fit of ratio $\phi^{\rm R}_{12}/(2\phi^{\rm R}_{13})$ and $\phi^{\rm R}_{13}$,
\item[fit 2:] conbined fit of eq.~\eqref{eq:pregolden} and $\phi^{\rm R}_{13}$.
\end{itemize}
Our results indicate that $\chi$PT suffers at our highest pion masses, which are around 420 MeV.
Therefore we introduce a cut in the pion mass at $m_{\pi} < 400$~MeV and $m_{\pi} < 300$~MeV to test
how much the results change. We take the results from fit 2 with a cut at $m_{\pi} < 400$~MeV
as our main result, and use the spread of central values in these two sets of fits to estimate the
systematics. The results have been crosschecked by various independent analyses.

\section{Results and outlook}

Our preliminary results for the strange and $u/d$ quark masses are
\begin{equation}
m_s^{\rm RGI} = 127.0(3.1)(3.2)\textrm{ MeV},\quad
m_{u/d}^{\rm RGI} = 4.70(15)(12)\textrm{ MeV}.
\end{equation}
The first error includes statistics/fitting and the second error is systematic. Using 4-loop
PT to convert RGI masses to $\overline{\rm MS}$ scheme at $\mu = 2$~GeV and $n_f=3$ gives
\begin{equation}
m_s^{\overline{\rm MS}} = 95.5(2.5)(2.4)\textrm{ MeV},\quad
m_{u/d}^{\overline{\rm MS}} = 3.53(12)(9)\textrm{ MeV}
\end{equation}
(the conversion factor is 1.330(13)). These results agree very well with the quark
masses listed in PDG~\cite{PDG}, $m_s^{\overline{\rm MS}} = 95^{+9}_{-3}$~MeV and
$m_{u/d}^{\overline{\rm MS}} = 3.5^{+0.5}_{-0.2}$~MeV. The agreement with other $n_f=2+1$
lattice results is also good: FLAG review \cite{FLAG} gives the lattice averages as
$m_s^{\overline{\rm MS}} = 92.0(2.1)$~MeV and $m_{u/d}^{\overline{\rm MS}} = 3.373(80)$~MeV.
Our result for the quark mass ratio is
\begin{equation}
\dfrac{m_s}{m_l} = 27.0(1.0)(0.4),
\end{equation}
compared to the PDG value~\cite{PDG} $m_s/m_l = 27.3(0.7)$ and the FLAG average~\cite{FLAG}
$m_s/m_l = 27.43(31)$.
Our results are still preliminary, and we will address the full error budget in a
forthcoming publication. The fairly large systematic uncertainties are due to the absence
of very chiral ensembles in this analysis, which could be improved on as further ensembles
become available.\\


{\bf Acknowledgements:}
The authors wish to thank P. Fritzsch, S. Schaefer and T. Korzec for their contributions and
numerous useful discussions.


\begin{thebibliography}{99}

\bibitem{Bruno:2016plf}
  M. Bruno, T. Korzec, and S. Schaefer [ALPHA Collaboration],
  \emph{Setting the scale for the CLS $2+1$ flavor ensembles},
  \emph{PRD} {\bf 95} (2017) 074504,
  [{\tt arXiv:1608.08900}].

\bibitem{Campos:2018ahf}
  I. Campos, P. Fritzsch, C. Pena, D. Preti, A. Ramos and A. Vladikas [ALPHA Collaboration],
  \emph{Non-perturbative quark mass renormalisation and running in $N_f=3$ QCD},
  \emph{EPJC} {\bf 78} (2018) 387,
  [{\tt arXiv:1802.05243}].

\bibitem{Bruno:2014jqa}
  M. Bruno {\it et al.} [ALPHA Collaboration],
  \emph{Simulation of QCD with $N_{f} = 2 + 1$ flavors of non-perturbatively improved Wilson fermions},
  \emph{JHEP} {\bf 1502} (2015) 043,
       [{\tt arXiv:1411.3982}].

\bibitem{Luscher:1984xn}
  M. L\"uscher and P. Weisz,
  \emph{On-Shell Improved Lattice Gauge Theories},
  \emph{Commun. Math. Phys.} {\bf 97} (1985) 59.
       [Erratum: \emph{Commun. Math. Phys.} {\bf 98} (1985) 433].

\bibitem{Bulava:2013cta}
  J. Bulava and S. Schaefer,
  \emph{Improvement of $N_f = 3$ lattice QCD with Wilson fermions and tree-level improved gauge action},
  \emph{Nucl. Phys.} {\bf B874} (2013) 188– 197,
       [{\tt arXiv:1304.7093}].

\bibitem{Luscher:2012av}
  M. L\"uscher and S. Schaefer,
  \emph{Lattice QCD with open boundary conditions and twisted-mass reweighting},
  \emph{Comput. Phys. Commun.} {\bf 184} (2013) 519–528,
       [{\tt arXiv:1206.2809}].  
       
\bibitem{Bulava:2015bxa}
  J. Bulava, M. Della Morte, J. Heitger, and C. Wittemeier [ALPHA Collaboration],
  \emph{Non-perturbative improvement of the axial current in $N_f = 3$ lattice QCD
    with Wilson fermions and tree-level improved gauge action},
  \emph{Nucl. Phys.} {\bf B896} (2015) 555–568,
       [{\tt arXiv:1502.04999}].

\bibitem{luscher:1996sc}
  M. L\"uscher, S. Sint, R. Sommer, and P. Weisz,
  \emph{Chiral symmetry and $\mathcal{O}(a)$ improvement in lattice QCD},
  \emph{ Nucl. Phys.} {\bf B478} (1996) 365–400,
       [{\tt arXiv:hep-lat/9605038}].
  
\bibitem{Bhattacharya:2005rb}
  T. Bhattacharya, R. Gupta, W. Lee, S. R. Sharpe, and J. M. Wu,
  \emph{Improved bilinears in lattice QCD with non-degenerate quarks},
  \emph{PRD} {\bf 73} (2006) 034504,
       [{\tt arXiv:hep-lat/0511014}].

 \bibitem{Luscher:2010iy}
   M. L\"uscher,
   \emph{Properties and uses of the Wilson flow in lattice QCD},
   \emph{JHEP} {\bf 08} (2010) 071,
        [{\tt arXiv:1006.4518}].
        [Erratum: JHEP03,092(2014)].

%
%

\bibitem{DallaBrida:2018tpn}
  M. Dalla Brida, T. Korzec, S. Sint, and P. Vilaseca,
  \emph{High precision renormalization of the flavour non-singlet Noether
    currents in lattice QCD with Wilson quarks},
  \emph{EPJC} {\bf 79} (2019) 23,
  [{\tt arXiv:1808.09236}].

  
\bibitem{Bar:2003mh}
  O. B\"{a}r, G. Rupak, and N. Shoresh,
  \emph{Chiral perturbation theory at $\mathcal{O}(a^2)$ for lattice QCD},
  \emph{PRD} {\bf 70} (2004) 034508,
       [{\tt arXiv:hep-lat/0306021}].
  
\bibitem{Herdoiza:2013sla}
  G. Herdo\'{\i}za, K. Jansen, C. Michael, K. Ottnad, and C. Urbach,
  \emph{Determination of Low-Energy Constants of Wilson Chiral Perturbation Theory},
  \emph{JHEP} {\bf 05} (2013) 038,
       [{\tt arXiv:1303.3516}].

\bibitem{CarlKoster} G. De Divitiis, P. Fritzsch, J. Heitger. C. C. K\"oster, S Kuberski,
  and A. Vladikas [ALPHA Collaboration] in preparation;
  we thank P. Fritzsch and C. C. K\"oster for providing us with their preliminary results.

\bibitem{Ramos}
  A. Ramos,
  \emph{Automatic differentiation for error analysis of Monte Carlo data},
  [{\tt arXiv:1809.01289}].

\bibitem{PDG}
  Particle Data Group Collaboration, M. Tanabashi \textit{et al.},
  \emph{Review of Particle Physics}, \emph{PRD} {\bf 98} (2018) 030001.

\bibitem{FLAG}
  S. Aoki \textit{et al.},
  \emph{Review of lattice results concerning low-energy particle physics},
  \emph{EPJ} {\bf C77} (2017) 112, [{\tt arXiv:1607.00299}].
  The results included in the averages are from:
  A. Bazavov \textit{et al.},
  \emph{MILC results for light pseudoscalars},
  \emph{PoS} {\bf CD09} (2009) 007, [{\texttt arXiv:0910.2966}];
  S. D\"urr \textit{et al.},
  \emph{Lattice QCD at the physical point: light quark masses},
  \emph{Phys.Lett.} {\bf B701} (2011) 265–268,
  [{\texttt arXiv:1011.2403}];
  S. D\"urr \textit{et al.},
  \emph{Lattice QCD at the physical point: simulation and analysis details},
  \emph{JHEP} {\bf 1108} (2011) 148, [{\texttt arXiv:1011.2711}];
  C. McNeile, C. T. H. Davies, E. Follana, K. Hornbostel and G. P. Lepage,
  \emph{High-precision c and b masses and QCD coupling from current-current
    correlators in lattice and continuum QCD},
  \emph{Phys. Rev.} {\bf D82} (2010) 034512, [{\texttt arXiv:1004.4285}];
  T. Blum {\textit et al.},
  \emph{Domain wall QCD with physical quark masses},
  \emph{Phys. Rev.} {\bf D93} (2016) 074505, [{\texttt arXiv:1411.7017}];
  A. Bazavov {\textit et al.},
  \emph{Staggered chiral perturbation theory in the two-flavor
    case and SU(2) analysis of the MILC data},
  \emph{PoS} {\bf LAT2010} (2010) 083, [{\texttt arXiv:1011.1792}];
  T. Burch {\textit et al.},
  \emph{Quarkonium mass splittings in three-flavor lattice QCD},
  \emph{Phys. Rev.} {\bf D81} (2010) 034508, [{\texttt arXiv:0912.2701}].
  
\end{thebibliography}
\end{document}